\documentclass[preprint,proceedings]{rmaa}

\suppressfulladdresses 

\usepackage{paralist}

\usepackage{psfrag,color}

\SetYear{2007}
\SetConfTitle{The Nuclear Region, Host Galaxy and Environment of Active Galaxies}

\title{Quasar Host Orientation and Polarization: Insights into the Type~1/Type~2 Dichotomy}

\author{
  B. Borguet,\altaffilmark{1,2}
  D. Hutsem\'{e}kers,\altaffilmark{1,3}
  G. Letawe,\altaffilmark{1}
  Y. Letawe,\altaffilmark{1}
  P. Magain,\altaffilmark{1}}

\altaffiltext{1}{Institut d'Astrophysique, 
  Universit\'{e} de Li\`{e}ge, All\'{e}e du 6 Ao\^{u}t 17, B-4000 Li\`{e}ge, Belgium (b.borguet@ulg.ac.be).}
\altaffiltext{2}{PhD. grant student F.N.R.S.}
\altaffiltext{3}{Senior Research associate F.N.R.S.}

\shortauthor{Borguet et al.}
\shorttitle{RevMexAA(SC) Insights into the Quasar Type~1/Type~2
Dichotomy}

\listofauthors{B. Borguet, D. Hutsem\'{e}kers, G. Letawe, Y.
Letawe \& P. Magain}
\indexauthor{Borguet, B.}
\indexauthor{Hutsem\'{e}kers, D.}
\indexauthor{Letawe, G.}
\indexauthor{Letawe, Y.}
\indexauthor{Magain, P.}

\abstract{We investigate correlations between the optical linear
polarization position angle and the orientation of the host galaxy/extended emission
of Type~1 and Type~2 Radio-Loud (RL) and Radio-Quiet (RQ) quasars. We have
used high resolution Hubble Space Telescope ({\it HST}) data and deconvolution
process to obtain a good determination of the host galaxy
orientation. With these new measurements and a compilation of data
from the literature, we find a significant correlation between the
polarization position angle and the position angle of the major axis of the host
galaxy/extended emission. The correlation appears different for Type~1 and Type~2 objects and depends on the
redshift of the source. Interpretations
in the framework of the unification model are discussed.}

\addkeyword{Quasars : general} \addkeyword{Polarization}

\begin{document}
\maketitle

\section{General}
\label{sec:intro}

There is a huge diversity of quasars in the Universe. In
particular, some of them harbor broad and narrow emission lines in their
spectrum (Type~1) while other ones only possess narrow
emission lines (Type~2). One can wonder whether the physical
processes at the origin of all quasars are the same.

In the intrinsically less powerful AGNs that are the Seyfert galaxies, the
discovery of broad emission lines in the polarized spectrum of
Type~2 objects provided strong support to the presence of a dusty 
torus oriented edge-on that blocks the direct view of a Type~1-like
central engine and broad emission line region (e.g. Antonucci \&
Miller 1985). A key question is whether this unification model (UM) also applies to quasars
since the presence of a dusty torus may be affected by the higher radiation
flux. The study of optical polarization is an interesting tool to
get some insights into the quasar inner structure.

The question we investigate relates to the possible existence of a
correlation between the linear optical polarization position angle
($\theta_{pola}$) and the orientation of the major axis of the host
galaxy/extended emission ($PA_{host}$) in the case of quasars. Such a study was
already undertaken by Berriman et~al. (1990), using ground based
data and showing a marginally significant correlation. Our
aim is to investigate the $\theta_{pola}/PA_{host}$ on the basis
of high resolution images of quasars (essentially from {\it HST} observations).

\section{Definition of the sample}
 \label{sec:sample}

  \subsection{Origin of data}

 Our initial sample was defined following one criterion: we searched
 for Type~1 and Type~2 quasars (RL and RQ but no blazars) for
 which high resolution visible/near-IR images (or derived
 host galaxy parameters) and optical polarization data
 were available in the literature. These samples are
 detailed in Borguet et~al. (2007).

   \subsection{Determination of $PA_{host}$}

 As several objects from this sample do not possess a $PA_{host}$ determined
 in the literature, we used the
 MCS deconvolution method (Magain et~al. 1998) to model and derive the host galaxy morphological
 parameters (the $PA_{host}$ and the ellipticity $a/b$).

 The image processing proceeds in two steps: first the proper subtraction of the bright central source from the images
 of (Type~1) quasars. This step generally allows the detection of the underlying host. The second step consists in the
 fitting of an analytical S\'{e}rsic galaxy profile properly convolved with the HST PSF to the point-source subtracted
 image. This process allows us to derive $PA_{host}$ and $b/a$ for
 a large part of the sample.

  \subsection{The polarimetric data}

 The polarimetric data mainly come from a compilation of polarization degree ($P$) and
 $\theta_{pola}$ measurements (Hutsem\'{e}kers et~al.
 2005). The $P$ and $\theta_{pola}$ measurements for the 2MASS and Type~2
 RQ quasars are respectively taken from Smith et~al. (2002) and
 Zakamska et~al. (2006).

 \section{The $PA_{host}-\theta_{pola}$ correlation}


 From the compilation, we selected only relevant and accurate
 data. Here are the criteria we used :
 \begin{asparaitem}
  \item We separated the quasars from the Seyfert galaxies by applying
  the $M_{V}\la-23$ criterion.
  \item{} We rejected objects with low ellipticity of the host galaxy ($b/a>0.9$) since
  the deduced $PA_{host}$ would have too large error bars.
  \item{} We considered objects for which significant
  polarization is detected, meaning objects such that $P/\sigma_{P} \ge 2$ (implying $\sigma_{\theta_{pola}} \le 14^{\circ}$).
 \end{asparaitem}

  For each quasar of the selected sub-sample, we
  computed the acute angle $\Delta \theta$ between the directions defined by the $PA_{host}$
  and $\theta_{pola}$ angles :
  $\Delta \theta = 90-|90-|\theta_{pola}-PA_{host}||$, $\Delta \theta \in [0^{\circ},90^{\circ}]$.
  In the following, we consider separately RQ and RL quasars as well as the visible and near-IR
  $PA_{host}$. 

  \subsection{The Radio-Quiet objects}
    \label{rqo}

\begin{figure}[!t]
\centering
  \includegraphics[width=\columnwidth]{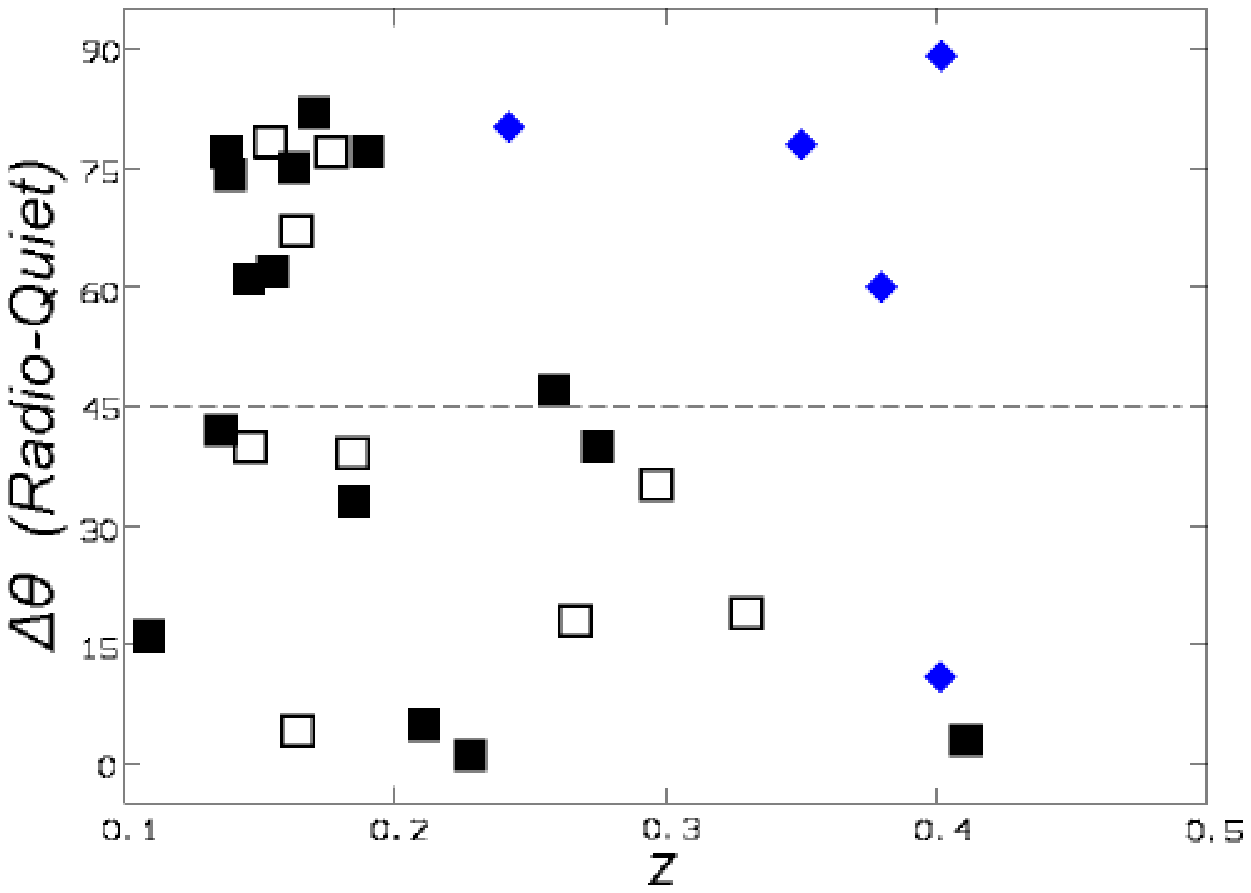}\\
  \includegraphics[width=\columnwidth]{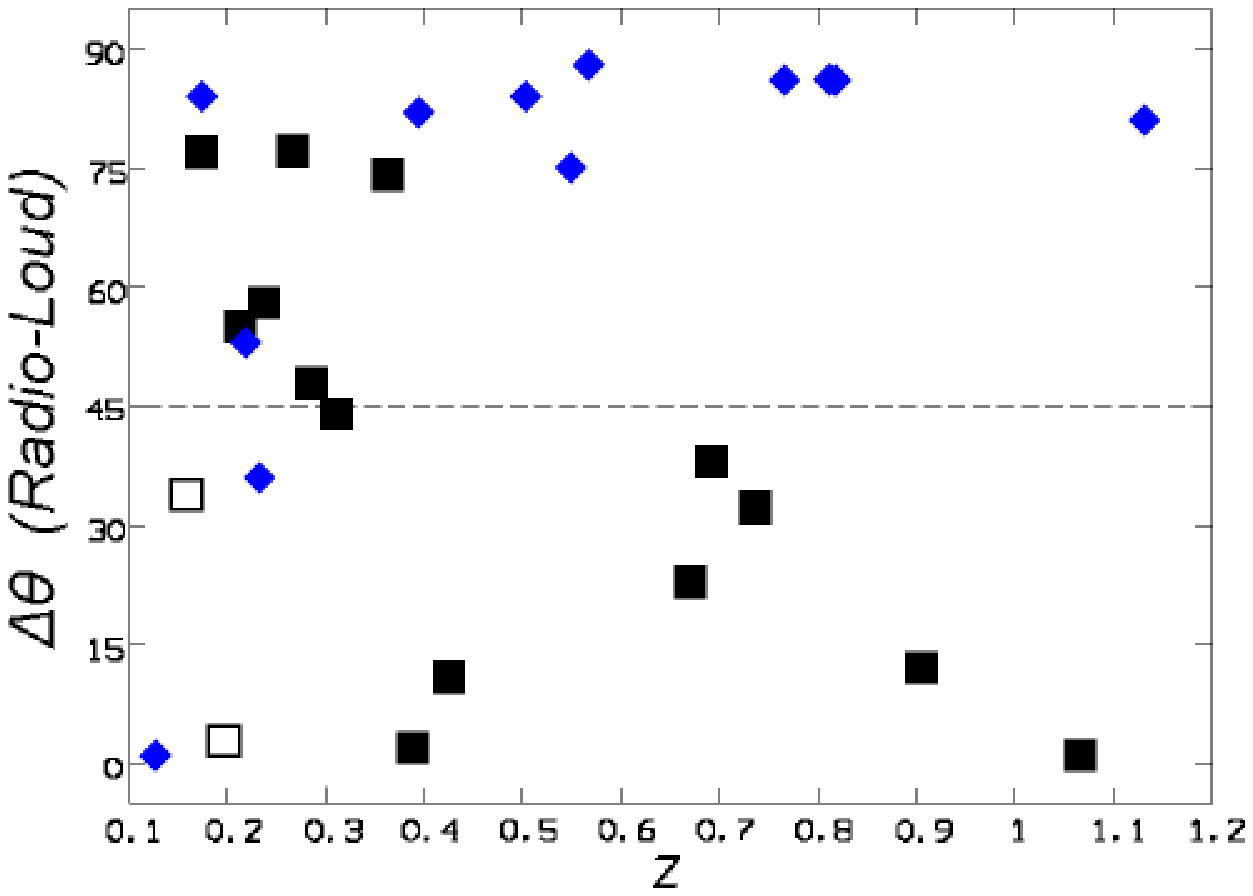}
  \caption{The acute angle $\Delta \theta$ between the polarization position
            angle $\theta_{pola}$ and the host galaxy position angle $PA_{host}$,
            as a function of the redshift $z$. The upper panel refers to Radio-Quiet objects while the lower
            panel refers to Radio-Loud objects. Type 1 objects are
            represented in both panel by squares and Type 2
            objects by diamonds. The filled symbols refer to
            objects with a polarization degree $P \geq 0.6 \%$}
              \label{fig:rqrldpa1z}%
\end{figure}

The behavior of the angle $\Delta \theta$ as a function of the redshift is illustrated in the upper panel of
Fig.~\ref{fig:rqrldpa1z} for both
Type~1 and Type~2 RQ quasars. While there seems to be no particular
behavior of $\Delta \theta$ at small redshifts ($z\la 0.2$), a
clear separation between Type~1 and Type~2 objects appears for
objects lying at higher z ($z\ga 0.2-0.3$). Indeed, for Type~1
quasars we observe that the polarization angle is preferentially
aligned with the host major axis (``alignment" meaning $\Delta
\theta < 45^{\circ}$), while for the Type~2 quasars an anti-alignment
($\Delta \theta > 45^{\circ}$) is observed (as already
noted by Zakamska et~al. 2006). The difference between the distribution of $\Delta \theta$ for Type~1 and Type~2
samples at $z>0.2$ is statistically significant (probability $\le 0.1\%$ with a 2 sample K-S test). 

 \subsection{The Radio-Loud objects}
    \label{rlo}

The lower panel of Fig.~\ref{fig:rqrldpa1z} summarizes the
$\Delta \theta$ behavior as a function of $z$ for RL quasars. Once
again, the distribution of $\Delta \theta$ for the high z objects is clearly
non-random. While we note that Type~1 quasars are systematically
found to lie at small offset angle ($\Delta \theta < 45^{\circ}$),
having their optical polarization preferentially parallel to their
host galaxy major axis, Type~2 quasars are found at higher offset
angles (as previously reported by Cimatti et~al. 1993). Again, a 
2 sample K-S test shows that there is a probability $\le 0.3\%$ 
that the Type~1 and Type~2 $\Delta \theta$ distributions come 
from the same parent sample.

 \subsection{Summary of the observations}
 \label{suma}

We can summarize our results as follows:
\begin{asparaitem}
\item{} While Type~2 quasars are known to exhibit an anti-alignment 
between the host/extended emission major axis
and the optical polarization, we find that an alignment dominates for Type~1 quasars.
\item{} This behavior seems to be independent of the
radio-loudness of the source.
\item{} We note a redshift dependence of the alignment effect. 
Moreover the $PA_{host}-\theta_{pola}$ correlation
using the $PA_{host}$ derived from near-IR images do not show
any particular behavior of the $\Delta \theta$ distribution.

\end{asparaitem}

\section{Discussion}
 
 The last observation of Sect.~\ref{suma} suggests that the alignment effect
 might be related to the rest-frame extended UV/blue part of the quasar emission as discussed below.
 
 \subsection{The extended UV/blue emission}

 Due to the absence of a bright point source, the morphology
 of Type~2 RL quasars has been extensively studied throughout the
 literature. An extended rest frame UV/blue emission has been observed in the
 optical images of the higher redshift objects ($z>0.5$) showing 
 a so-called ``alignment effect" with the radio jet,
 both being preferentially co-linear (McCarthy et~al. 1987, Chambers et~al.
 1987). Using polarimetric measurements of such targets, Cimatti et~al. (1993) noted
 an anti-alignment between the directions defined by the polarization angle
 and the extended UV/blue emission. This observation
 suggested that at least part of the UV/blue light might be related to scattering in
 extended polar regions.

 Due to their apparent faintness and absence of strong radio
 counterpart, Type~2 RQ quasars have long been searched for. The recent imaging
 and spectropolarimetric observations of these objects revealed, as 
 in the case of Type~2 RL quasars, the presence of an extended UV/blue 
 light, whose extension was found to be anti-aligned with the
 polarization angle, arguing in favor of a scattering origin of the blue light (Zakamska et~al. 2006).

 In the case of Type~1 quasars, the detection of an
 hypothetical UV/blue extension is hindered by the central source,
 whose contribution remains difficult to properly subtract.
 However, if the UM applies to quasars, such an extended polar scattering
 region may also be present in Type~1 objects but might lead to
 different polarization properties given the smaller viewing angle
 to the system.

 \subsection{Interpretation in the framework of the UM}
 
 Seyfert galaxies are known to exhibit a kind of ``alignment effect" between their optical polarization and
 radio-jet position angle (Smith et~al. 2002 and references therein). In order to describe their observations, Smith et~al.~introduced a two component
 scattering model in which the polarization is produced in two separate scattering 
 regions. Polar regions produce a polarization direction anti-aligned with the
 symmetry axis of the accretion disk while an equatorial region located inside the torus produces a
 polarization aligned with the system axis.
 
 Our observations fit this two component model. In Type~2 quasars (either RL or RQ)
 the observed polarization is only produced by the polar region (the equatorial one being masked by the
 obscuring material). In Type~1
 quasars, the higher symmetry of the polar region let the equatorial
 region dominate the polarized flux, resulting in a polarization
 angle parallel to the polar extended UV/blue emission.

 \section{Conclusions}

We can summarize our results as follow : 
\begin{asparaitem}
\item{} While Type~2 RL and RQ quasars are known to exhibit an anti-alignment between the major axis of hteir host/extended emission and their optical polarization, we find that an alignment is
mostly observed for Type~1 quasars.
\item{} The redshift dependence of the alignment effect, and
lack of correlation with the near-IR
$PA_{host}$, suggest that it might be related to the rest-frame extended
UV/blue emission of quasars.
\item{}We show that these observations
can be interpreted in the framework of the unification model + a two component
scattering model.
\end{asparaitem}

\end{document}